\documentclass[twocolumn,prl,floatfix]{revtex4}
\usepackage{amssymb,amsmath,latexsym}
\usepackage{bm}
\usepackage[dvips]{epsfig}
\usepackage{color}

\renewcommand{\vec}[1]{\bm{\mathrm{#1}}}
\newcommand{\vhat}[1]{\hat{\bm{\mathrm{#1}}}}

\begin{document}

%%%%%%%%%%%%%%%%%%%%%%%%%%%%%%%%%%%%%%%%%%%%%%%%%%%%%%%%%%%%%%%%%%%%
\title{Magnetization dynamics induced by in-plane currents
in ultrathin magnetic nanostructures}
\author{Kyoung-Whan Kim$^1$}
\author{Soo-Man Seo$^2$}
\author{Jisu Ryu$^1$}
\author{Kyung-Jin Lee$^{2,3,4}$}\email{kj_lee@korea.ac.kr}
\author{Hyun-Woo Lee$^1$}\email{hwl@postech.ac.kr}
\affiliation{$^1$PCTP and Department of Physics, Pohang University
of Science and Technology, Pohang, 790-784, Korea\\$^2$Department of
Materials Science and Engineering, Korea University, Seoul 136-701,
Korea\\$^3$Center for Nanoscale Science and Technology, National Institute of Standards and Technology, Gaithersburg, Maryland 20899-8412, USA\\$^4$Maryland Nanocenter, University of Maryland, College Park, MD 20742, USA}

\date{\today}

%%%%%%%%%%%%%%%%%%%%%%%%%%%%%%%%%%%%%%%%%%%%%%%%%%%%%%%%%%%%%%%%%%%%%

\begin{abstract}
Ultrathin magnetic systems have properties qualitatively different
from their thicker counterparts, implying that different physics
governs their properties. We demonstrate that various such
properties can be explained naturally by the Rashba spin-orbit
coupling in ultrathin magnetic systems. This work will be valuable
for the development of next generation spintronic devices based on
ultrathin magnetic systems.
\end{abstract}


\maketitle
%%%%%%%%%%%%%%%%%%%%%%%%%%%%%%%%%%%%%%%%%%%%%%%%%%%%%%%%%%%%%%%%%%%%%%%%%%

Electric control of magnetic systems carries high potential towards device
applications~\cite{Chappert07NM,Katine08JMMM} such as magnetic memory and
logic. Spin-transfer torque (STT)~\cite{Slonczewski96JMMM,Berger96PRB} is an
efficient way to achieve the electric control of magnetic nanostructures.
% and is the core
%mechanism behind the industrial attempt to realize post-silicon
%era by using MgO-based magnetic tunnel junctions (MTJs).
%
In view of device applications, magnetic nanostructures such MgO-based
magnetic tunnel junctions are superior to silicon-based nanostructures in the
simultaneous realization of nonvolatility and speed, but are estimated to
require  100 times more energy~\cite{Amiri11IEEE} than silicon-based CMOS
devices to write an information bit. This energy cost problem limits scope of
device applications based on magnetic nanostructures.
Since the writing energy decreases as a magnetic layer in magnetic
nanostructures becomes thinner~\cite{Katine08JMMM}, properties of
ultrathin magnetic layers are under intense
investigation~\cite{Ikeda11NatMat}.

\begin{figure}
\includegraphics{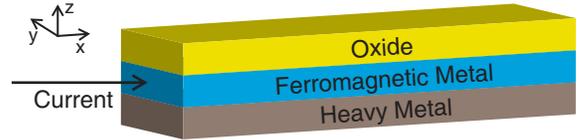}
\caption{(color online) Schematic structure of ultrathin magnetic
nanostructure, where an ultrathin ($\approx$ 1 nm) ferromagnetic
layer is sandwiched between a heavy metallic layer and an
insulating oxide layer. Examples include Pt/Co/AlO$_x$ and
Ta/CoFeB/MgO. }
\label{Fig:thin-layer}
\end{figure}

While the magnetization switching for the information writing is
conventionally achieved by a current perpendicular to a magnetic
layer, a recent experiment~\cite{Miron11Nat} found that an {\it
in-plane} current can also switch the uniform magnetization of an
ultrathin ($\approx$ 1 nm) magnetic layer (Co)  sandwiched between
a heavy metal layer (Pt) and an oxide layer (AlO$_x$)
(Fig.~\ref{Fig:thin-layer}).
Since the cross-sectional area (in $yz$-plane) for the in-plane
current can be orders of magnitude smaller than the
cross-sectional area (in $xy$-plane) for the perpendicular
current, this alternative switching scheme may reduce the current
required for the switching and the switching energy.
%the writing energy may be reduced provided that the threshold
%current density for the switching remains similar in the two
%current geometries.
%
It was also reported that the magnetic domain wall (DW) in the ultrathin
magnetic layer moves as fast as 400 m/s~\cite{Miron11NM} when in-plane
current is supplied. This velocity is about 4 times higher than the highest
velocity reported for thicker magnetic layers~\cite{Hayashi07PRL}.
Thus the in-plane current effects on ultrathin magnetic systems
open an attractive alternative path towards powerful spintronic
devices.

Ultrathin magnetic systems are interesting in view of fundamental science as
well. Various features of ultrathin magnetic systems cannot be explained by
existing theoretical knowledge learned from measurements on thicker
counterparts; for example (i) magnetization switching by  in-plane
current~\cite{Miron11Nat} instead of  perpendicular current, (ii)  DW motion
against the electron flow direction~\cite{Moore08APL,Miron11NM} instead of
along it, and (iii)  anomalously high DW speeds~\cite{Miron11NM}. These
anomalies imply that ultrathin magnetic systems are not mere thin limits of
thicker counterparts but are qualitatively different systems governed by
different physics. A clear understanding of their core physics will be highly
valuable for developments of spintronic devices.

%Experiments provide a hint to the core physics; The anomalies
%disappear when the ultrathin magnetic layer is put into a
%symmetric environment, for instance, by replacing the oxide layer
%with another heavy metal layer made of the same metallic element
%as in the bottom metallic layer (Fig.~\ref{Fig:thin-layer}). Thus
%the structural asymmetry is important.
%%%%%%%%%%%%%%%%%%%%%%%%%%%%%%%%%%%%%%%%%
%%%%%%%%%%%%%%%%%%%%%%%%%%%%%%%%%%%%%%%%%
Since the upper and lower layers of the ultrathin magnetic layer are made of
quite different materials (Fig.~\ref{Fig:thin-layer}), the ultrathin magnetic
layer has the structural asymmetry. Here we demonstrate theoretically that if
the Rashba spin-orbit coupling (RSOC)~\cite{Bychkov84JETPL} due to the
asymmetry is sufficiently strong, all the anomalies (i), (ii), and (iii) can
be explained naturally. To be specific, we show that the in-plane current
gives rise to a torque proportional to $\alpha_{\rm R}
 \vec{m}\times[ \vec{m}\times (\vhat{z}\times\vec{j}_e  )]$~\cite{comment-Arne},
 where $\alpha_{\rm R}$ is the parameter describing the strength of the RSOC,
 $\vec{j}_e$ is the in-plane current density in the ultrathin magnetic layer,
$\vhat{z}$ is the unit vector perpendicular to the layer, and
$\vec{m}$ is the unit vector along the magnetization in the
ultrathin magnetic layer. Since this torque has the same form as
the Slonczewski STT~\cite{Slonczewski96JMMM} in magnetic
multilayers, we call it Slonczewski-like STT (SL-STT). We
demonstrate that the SL-STT explains all three anomalies
naturally. Recently an experiment~\cite{Liu11preprint} proposed
that the anomaly (i) may arise due to the spin Hall effect (SHE)
in the heavy metal layer. Here we demonstrate that the SHE alone
cannot explain the anomalies (ii) and (iii).

First, we demonstrate that RSOC generates the SL-STT. Conduction electrons in the ultrathin ferromagnetic layer can be described
by the Hamiltonian,
\begin{equation}
H=\frac{\vec{p}_{\rm
op}^2}{2m_e}+J_{\rm ex}\vec{\sigma}\cdot\vec{m}+\frac{\alpha_{\rm R}}{\hbar}(\vec{\sigma}\times\vec{p}_{\rm
op})\cdot\vhat{z}+H_{\rm rel}, \label{Eq:Hamiltonian}
\end{equation}
where $\vec{p}_{\rm op}$ is the momentum operator, $m_e$ is the
electron mass, $J_{\rm ex}$ $(>0)$ is the exchange energy, and
$\vec{\sigma}$ is the Pauli matrix.
% and $\vhat{z}$ is the normal vector to the ultrathin ferromagnetic layer.
%Here, $J_{ex}$ is positive since the spin
%direction is antiparallel to the magnetic moment in general.
The third term in Eq.~(\ref{Eq:Hamiltonian}) is the RSOC
Hamiltonian and the last term describes spin relaxation processes
such as electron scattering.
When expressed in terms of the kinematic velocity operator
%$\vec{\pi}_{\rm op}\equiv m_e\vec{v}_{\rm op}$, where
$\vec{v}_{\rm op}\equiv \vec{p}_{\rm op}/m_{e}+(\alpha_{\rm
R}/\hbar)\vhat{z}\times\vec{\sigma}$,
%[\vec{r},H]/i\hbar$ is the velocity operator,
%It is convenient to rewrite down Eq. (\ref{Eq:Hamiltonian}) . Dropping $c$-number,
Eq. (\ref{Eq:Hamiltonian}) becomes $H=m_e\vec{v}_{\rm
op}^2/2+J_{\rm ex}\vec{\sigma}\cdot\vec{m}+H_{\rm rel}$, where a
trivial $c$-number has been neglected.
%
%To make a further progress, we recall that the exchange energy
%$J_{\rm ex}$ is the largest energy scale affecting the spin
%dynamics in conventional metallic ferromagnets (such as Co or Fe).
Due to the strong exchange energy $J_{\rm ex}$, which is the
largest energy scale affecting the spin dynamics in conventional
metallic ferromagnets (such as Co or Fe), electrons in energy
bands still carry largely majority and minority spin directions.
However within each of these bands, the individual spins and the
net spin are not nessarilily collinear with the magnetization due
to tilting of the spins by RSOC. In the following, we treat the
electrons on the majority and minority sheets of the Fermi surface
separately, but need to treat each spin density as a vector
$\vec{s}_\pm\equiv \langle \vec{\sigma} \rangle_\pm$, since it is
not aligned with the magnetization. Here
$\langle\cdots\rangle_\pm$ denotes local average over
majority/minority electrons.
%
%The spin continuity equation becomes
%%\begin{equation}
%%\frac{\partial\vec{\sigma}}{\partial t}
%$\partial\vec{\sigma}/\partial t +\nabla\cdot\mathcal{J}_{\rm
%OP}=
%%\frac{1}{i\hbar}[\sigma,H]
%[J_{ex}\vec{\sigma}\cdot\vec{m}+\alpha_R\hbar^{-1}(\vec{\sigma}\times\vec{p}_{\rm
%op})\cdot\vhat{z}]/(i\hbar) + \vec{\Gamma},
%%(\vec{\sigma})
%%,\label{Eq:continuity equation}
%%\end{equation}
%$
%
%where $\vec{\Gamma}=[\sigma,H_{\rm rel}]/(i\hbar)$ and
%$\mathcal{J}_{OP}$ is the spin-current tensor operator.
%
The spin continuity equation determined by $H$ allows one to
derive the Bloch equation for $\vec{s}_\pm$,
\begin{equation}
\frac{\partial\vec{s}_\pm}{\partial t}+\nabla\cdot\mathcal{J}_\pm
= -\frac{\vec{s}_\pm}{\tau_{\rm ex}}\times\left[\vec{m}
+\frac{2\alpha_{\rm R}m_e\tau_{\rm
ex}}{\hbar^2}(\vec{v}_\pm\times\vhat{z})\right]
+\langle\vec{\Gamma}\rangle_\pm,\label{Eq:continuity equation2}
\end{equation}
where  $\tau_{\rm ex}\equiv\hbar/2J_{\rm ex}$,
% is the characteristic time of the exchange coupling,
$\vec{v}_\pm=\langle\vec{v}_{\rm op}\rangle_\pm$, and
$\vec{\Gamma}=[\vec{\sigma},H_{\rm rel}]/i\hbar$, and the
spin-current tensor density
%$\mathcal{J}_\pm=\langle\mathcal{J}_{\rm op}\rangle_\pm$.
$\mathcal{J}_\pm= \vec{v}_\pm\otimes\vec{s}_\pm$. The approximation $\langle (\vec{\sigma})_i (\vec{v}_{\rm op})_j
+(\vec{v}_{\rm op})_j(\vec{\sigma})_i \rangle_\pm=2(\vec{s}_\pm)_i
(\vec{v}_\pm)_j$ is used in Eq.~(\ref{Eq:continuity
equation2})~\cite{comment:anomalous velocity correction}.
Recalling that STT is determined by the transverse component
$\delta\vec{s}_\pm$ of $\vec{s}_\pm$ perpendicular to $\vec{m}$,
it is useful to separate $\vec{s}_\pm$ into longitudinal and
transverse components, $\vec{s}_\pm = \mp n_\pm
\vec{m}+\delta\vec{s}_\pm$, where $n_\pm=\mp \vec{s}_\pm \cdot
\vec{m}$ is the longitudinal spin accumulation.
%The first term in right-hand side results from nonvanishing commutator between
%$\vec{\sigma}$ and the anomalous velocity correction in
%$\vec{p}_{\rm op}'$.
%In the definitions of $\vec{s}_\ud$ and
%$\mathcal{J}_\ud$, minus signs are introduced since the electron
%spin is antiparallel to its magnetic moment.
%
One then makes the relaxation time
approximation~\cite{Zhang04PRL},
$\langle\vec{\Gamma}\rangle_\pm=-\delta\vec{s}_\pm/\tau_{\rm sr}$,
where $\tau_{\rm sr}$ is the transverse spin relaxation time.
%and $\delta\vec{s}_\ud$ is the
% component of $\vec{s}_\ud$ perpendicular to
%$\vec{m}$.
In this approximation, the relaxation of the longitudinal spin
component is neglected since the transverse relaxation is much
faster in conventional metallic ferromagnets and also the
longitudinal spin component does not affect the STT.
%
%After some algebra~\cite{comment:algebra}, one obtains
%\begin{equation}
%D_t^{\pm\hspace{-1pt}}\vec{s}_\pm =-\frac{2\alpha_{\rm
%R}m_e}{\hbar^2}\vec{s}_\pm\times(\vec{v}_\pm\times\vhat{z})
%-\frac{1}{\tau_{\rm
%ex}}[\vec{s}_\pm-\beta(\vec{s}_\pm\times\vec{m})]\times\vec{m},\label{Eq:model
%Eq}
%\end{equation}
%where $D_t^{\ud\hspace{-1pt}}=\partial_t+\vec{v}_\ud\cdot\nabla$ and
%$\beta=\tau_{\rm ex}/\tau_{\rm sr}$.
%
%When the exchange energy $J_{ex}$ is the largest energy scale
%affecting the spin dynamics, one can express the solution of
%Eq.~(\ref{Eq:model Eq}) as a perturbative series with $\tau_{ex}$
%as an expansion parameter.
%
When $\vec{v}_\pm$ is assumed to be
homogeneous~\cite{comment:algebra} within the magnetic layer, one
can determine from Eq.~(\ref{Eq:continuity equation2}) the
following approximate solution for $\delta \vec{s}_\pm$,
\begin{equation}
\delta\vec{s}_\pm=\pm n_\pm\tau_{\rm
ex}\frac{(\beta+\vec{m}\times)}{1+\beta^2}
\left[D_t^{\pm\hspace{-1pt}}\vec{m} +\frac{2\alpha_{\rm
R}m_e}{\hbar^2}\vec{m}\times(\vec{v}_\pm\times\vhat{z})\right],
\end{equation}
where $\beta=\tau_{\rm ex}/\tau_{\rm sr}$ and
$D_t^{\pm\hspace{-1pt}}=\partial_t+\vec{v}_\pm\cdot\nabla$.
Corrections to this solution are of higher order in $\tau_{\rm
ex}$ and may be neglected in the strong $J_{\rm ex}$ limit.
Finally from the relation $\vec{T}=\mu_B\tau_{\rm
ex}^{-1}\vec{m}\times\delta\vec{s}$ between STT $\vec{T}$ and the
total transverse spin density $\delta \vec{s}=\delta\vec{s}_+ +
\delta\vec{s}_-$, one obtains
%the total STT $\vec{T}$,
%\begin{equation}
%\vec{T}=-\frac{n_s(1-\beta\vec{m}\times)}{1+\beta^2}
%\left[D_t\vec{m}+\frac{2\alpha_R}{\hbar^2}\vec{m}\times(\vec{p}\times\vhat{z})\right].\label{Eq:STT}
%\end{equation}
%When this result is inserted into the Landau-Lifshitz-Gilbert
%(LLG) equation
%\begin{equation}
the Landau-Lifshitz-Gilbert equation $\partial\vec{M}/\partial
t=-\gamma_0\vec{M}\times\vec{H}_{\rm eff}+(\alpha_0/M_{\rm
s})\vec{M}\times\partial\vec{M}/\partial t+\vec{T}$, where
$\vec{H}_{\rm eff}$ is a sum of an external magnetic field and
effective magnetic fields due to magnetic anisotropy and magnetic
exchange energy. $\vec{M}=M_{\rm s}\vec{m}$ is the magnetization
and $M_{\rm s}$ is saturation magnetization in the ultrathin
magnetic layer. After grouping together terms of the same
structure, one obtains
%\end{equation}
%one obtains the modified LLG equation~
\begin{eqnarray}
\frac{\partial\vec{M}}{\partial t}&=&-\gamma\vec{M}\times
\left(\vec{H}_{\rm eff}+\vec{H}_{\rm R}-\frac{\beta}{M_{\rm s}}\vec{M}\times\vec{H}_{\rm R}\right)\nonumber\\
&&+\frac{\alpha}{M_{\rm s}}\vec{M}\times\frac{\partial\vec{M}}{\partial t}
+\frac{\mu_{\rm B}P}{eM_{\rm s}(1+\beta^2)}(\vec{j}_e\cdot\nabla)\vec{M}\nonumber\\
&&-\frac{\beta\mu_{\rm B}P}{eM_{\rm s}^2(1+\beta^2)}
\vec{M}\times(\vec{j}_e\cdot\nabla)\vec{M}.\label{Eq:LLG-modified}\\
\vec{H}_{\rm R}&=&\frac{\alpha_{\rm R}m_eP}{\hbar
eM_{\rm s}(1+\beta^2)}(\vhat{z}\times\vec{j}_e),
\end{eqnarray}
where $\vec{H}_{\rm R}$ is the additional effective magnetic field
due to RSOC,
%$\vec{j}_e$ is the charge current density,
$\vec{j}_e=-e(n_+ \vec{v}_+ + n_- \vec{v}_-)$, and
$P\vec{j}_e=-e(n_+ \vec{v}_+ - n_- \vec{v}_-)$.
Note that the first and fourth terms on the right-hand-side of
Eq.~(\ref{Eq:LLG-modified}) contain the renormalized gyromagnetic
ratio $\gamma$ and the renormalized Gilbert damping $\alpha$ given
by $\gamma_0/\gamma=1+(n_+ - n_-)/[M_{\rm s}(1+\beta^2)]$, and
$\gamma\alpha/\gamma_0=\alpha_0+\beta(n_+-n_-)/[M_{\rm
s}(1+\beta^2)]$.
The last and second to the last terms are the nonadiabatic
STT~\cite{Zhang04PRL,Thiaville05EL} and the adiabatic
STT~\cite{Tserkovnyak08JMMM}. These four terms govern the
magnetization dynamics in thicker magnetic systems.

The second and third terms are additional STTs due to RSOC and may have
sizable magnitude only in ultrathin magnetic systems since $\alpha_{\rm R}$
decays as the magnetic layer becomes thicker. The second term
$-\gamma\vec{M}\times \vec{H}_{\rm R}$ has the same structure as the first
term and thus we call it field-like STT (FL-STT). The FL-STT was derived
previously~\cite{Obata08PRB,Manchon08PRB} and Ref.~\cite{Miron10NM} reported
its experimental confirmation. The FL-STT alone, however, can explain none of
the anomalies, as we demonstrated recently~\cite{Ryu11}. The third term is
the very SL-STT that this calculation aims to derive.

Next we explain the anomalies in terms of the SL-STT. Explanation
for the anomaly (i) is trivial. Recalling that the SL-STT due to
the in-plane current has the exactly same structure as the
Slonczewski STT~\cite{Slonczewski96JMMM} due to the perpendicular
current in metallic spin valve systems and  that the Slonczewski
STT induces the magnetization switching in the spin valve systems,
it is easy to understand that the in-plane current can induce the
magnetization switching through the action of the SL-STT.

%%%%%%%%%%%%%%%%%%%%%%%%%%%%%%%%%%%%%%%%%
\begin{figure}
\includegraphics{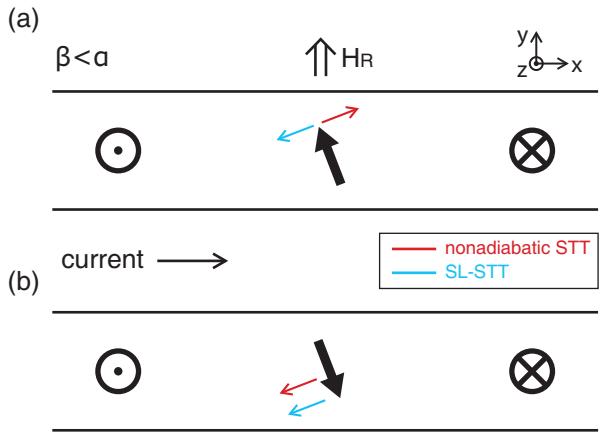}
\caption{(color online) Two possible structures of the Bloch DW
when the current flows to the right and
$\alpha>\beta$~\cite{addenda}. Thick and thin arrows represent the
directions of $\vec{m}$ and STTs, respectively. Note that the
direction of $\vec{m}$ deviates from $\pm \vhat{y}$, which is a
generic feature of a moving DW~\cite{Zhang04PRL,Thiaville05EL}
with $\alpha\neq \beta$.}
\label{Fig:Bloch DW}
\end{figure}
%%%%%%%%%%%%%%%%%%%%%%%%%%%%%%%%%%%%%%%%%

The anomalies (ii) and (iii) are less trivial to explain since
they arise from the combined action of the SL- and FL-STTs.
To explain the anomalies, we begin with two possible structures of
the Bloch DW (Fig.~\ref{Fig:Bloch DW}) in ultrathin magnetic
systems (such as Pt/Co/AlO$_x$~\cite{Miron10NM,Miron11NM}) with
the perpendicular magnetic anisotropy.
In the absence of RSOC and when $\vec{j}_e$ is not large enough to cause the
DW structural instability (Walker breakdown)~\cite{Thiaville05EL}, the two
structures are equivalent  in term of both stability and dynamics (same DW
velocity $v_{\rm DW}$).
The first effect of RSOC is to break the dynamic equivalence; at
the DW center in Fig.~\ref{Fig:Bloch DW}(a)/(b), the SL-STT is
anti-parallel/parallel to the nonadiabatic STT, effectively
cancelling/enlarging the effect of the nonadiabatic STT. Recalling
that the nonadiabatic STT determines $v_{\rm
DW}$~\cite{Zhang04PRL,Thiaville05EL}, this implies that the SL-STT
reduces/increases $v_{\rm DW}$.
When RSOC is sufficiently strong, it is even possible that the
SL-STT over-cancels the nonadiabatic STT in Fig.~\ref{Fig:Bloch
DW}(a), so that
%the the sum of the SL-STT and
%the nonadiabatic STT {\it reverses} its direction. In such a
%situation,
$v_{\rm DW}$ reverses its sign.
When RSOC is even stronger, $v_{\rm DW}$ will be large with the
reversed sign, implying that the DW in Fig.~\ref{Fig:Bloch DW}(a)
moves {\it fast against} the electron flow direction.
By the way, the FL-STT does not affect $v_{\rm DW}$~\cite{Ryu11}
since it is perpendicular to the nonadiabatic STT at the DW
center.
The second effect of RSOC is to break the stability equivalence.
The effect of the FL-STT on the stability can be understood from
the direction of the effective field $\vec{H}_{\rm R}$. Since the
effective energy density $-\vec{H}_{\rm R}\cdot\vec{M}$ is
negative/positive at the DW center for the DW structure in
Fig.~\ref{Fig:Bloch DW}(a)/(b), the FL-STT makes the DW structure
in Fig.~\ref{Fig:Bloch DW}(a) more stable than the other
structure. Moreover when $\vec{H}_{\rm R}$ is sufficiently strong,
the DW structure in Fig.~\ref{Fig:Bloch DW}(b) becomes unstable
and evolves to the stable DW structure in Fig.~\ref{Fig:Bloch
DW}(a)~\cite{Ryu11}.  By the way, the SL-STT has much weaker
effect on the stability than the FL-STT since, according to
Eq.~(\ref{Eq:LLG-modified}), the SL-STT is smaller than the FL-STT
in magnitude by factor $\beta$ and this nonadiabaticity parameter
$\beta$ is known~\cite{Zhang04PRL} to be smaller than 1.
%. Recalling that $\beta$ represents
%the relative magnitude of the nonadiabatic STT with respect to the
%adiabatic STT and that the nonadibatic STT is smaller than the
%adiabatic STT, one finds that the RSOC effect on the stability is
%governed by the FL-STT.
%
Then combining the above information, we find that there is only
one stable DW structure [Fig.~\ref{Fig:Bloch DW}(a)] when RSOC is
sufficiently strong and that it moves fast against the electron
flow direciton, explaining both the anomalies (ii) and (iii).

The experiment~\cite{Miron11NM}, which reported the anomalies (ii)
and (iii), provides sufficient information to test this
explanation. According to Eq.~(\ref{Eq:LLG-modified}), the
magnitude of the SL-STT is $\gamma \beta |\vec{M}\times
\vec{H}_{\rm R}|= \gamma \beta M_{\rm s} |\vec{H}_{\rm R}|\sin
\phi$, where $\phi$ is the DW tilting angle between $+\vhat{y}$
and $\vec{M}$ at the DW center, and the experiment reported
$|\vec{H}_{\rm R}|\sim$ 1T at $|\vec{j}_e|=10^{12}$ A/m$^2$. On
the other hand, the magnitude of the nonadiabatic STT is $\beta
b_j |\partial \vec{M}/\partial x|\sim \beta b_j M_{\rm s}/
\lambda$, where  the DW width $\lambda$ is about 5 nm and
$b_j=\mu_{\rm B} P j_e / [eM_{\rm s} (1+\beta^2)]$ is of the order
of 100 m/s at $|\vec{j}_e|=10^{12}$ A/m$^2$. Thus the relative
magnitude of the SL-STT with respect to the nonadiabatic STT is
roughly given by $\tilde{\alpha}_{\rm R}\sin \phi$, where the
dimensionless parameter $\tilde{\alpha}_{\rm R}\equiv
\pi\alpha_{\rm R} m_e\lambda/\hbar^2 = (\pi/2) (\gamma\beta M_{\rm
s} |\vec{H}_{\rm R}|)/(\beta b_j M_{\rm s}/ \lambda)$ is of the
order of 10. Thus unless $\phi$ is too small, the SL-STT can be
indeed larger than the nonadiabatic STT and reverse the sign of
$v_{\rm DW}$.

%%%%%%%%%%%%%%%%%%%%%%%%%%%%%%%%%%%%%%%%%
\begin{figure}
\includegraphics{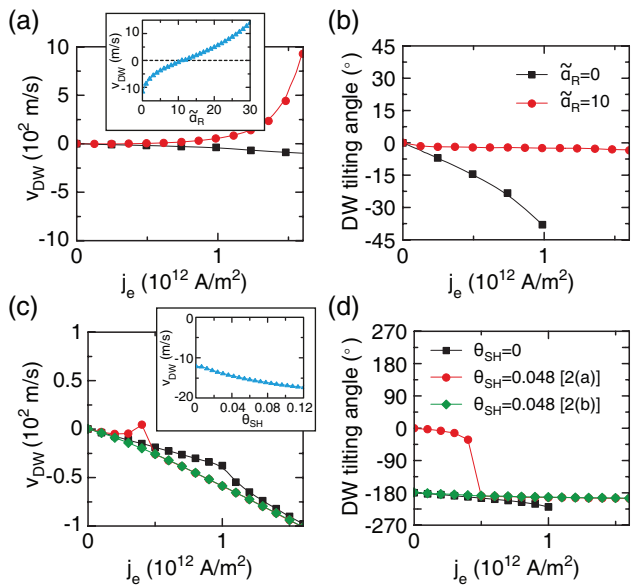}
\caption{(color online) Micromagnetic simulation results of the
current-driven DW motion. In (a) and (b), RSOC effects are
examined by using Eq.~(\ref{Eq:LLG-modified}).
$v_{\rm DW}$ (a) and the DW tilting angle $\phi$ (b) [measured
clockwise from $+\vhat{y}$ direction in Fig.~\ref{Fig:Bloch
DW}(a)] as a function of $j_e$ for $\tilde{\alpha}_R=0$ (black
squares)  and 10 (red circles).
As shown in (b), the Walker breakdown is suppressed when
$\tilde{\alpha}_{\rm R}=10$ whereas it occurs for $j_e> 1.0\times
10^{12}$ A/m$^2$ when $\tilde{\alpha}_{\rm R}=0$.
The inset in (a) shows $v_{\rm DW}$ as a function of
$\tilde{\alpha}_{\rm R}$ for $j_e=3.0\times 10^{11}$ A/m$^2$.
All results in (a) and (b) are for the stable DW configuration in
Fig.~\ref{Fig:Bloch DW}(a).
In (c) and (d), SHE effects are examined by setting $\vec{H}_{\rm
R}=0$ and instead adding to Eq.~(\ref{Eq:LLG-modified}) the
SHE-induced SL-STT
$\gamma \vec{M} \times [ (\theta_{\rm SH}\vec{M}/{M_{\rm
s}})\times ( H_{\rm S} \vhat{y} ) ]$,
where
%$\theta_{\rm SH}$ is the spin Hall angle,
$H_{\rm S}= \hbar j_{e,{\rm N}}/(2eM_{\rm s} t_{\rm F})$, $t_{\rm
F}$ is the thickness of the ultrathin magnetic layer, and
$j_{e,{\rm N}}$ is the charge current density in the heavy metal
layer. $j_{e,{\rm N}}=j_e$ is assumed.
$v_{\rm DW}$ (c) and $\phi$ (d) as a function of $j_e$ for
$\theta_{\rm SH}=0$ (black squares) and 0.048 [red circles and
green diamonds for the DW structures in Figs.~\ref{Fig:Bloch
DW}(a) and (b)]. Red circles are not visible for $j_e>5\times
10^{11}$ A/m$^2$ since they overlap with green diamonds.
%
%When $\theta_{\rm SH}=$-0.048, the sudden jump in (d) indicates
%that the DW configuration in Fig.~\ref{Fig:Bloch DW}(a) becomes
%unstable for $j_e>0.5\times 10^{12}$ A/m$^2$ and makes a chiral
%switching to the DW configuration in Fig.~\ref{Fig:Bloch DW}(b).
%
The inset in (c) shows $v_{\rm DW}$ as a function of $\theta_{\rm
SH}$ at $j_e=3.0\times 10^{11}$ A/m$^2$ for the stable DW
configuration in Fig.~\ref{Fig:Bloch DW}(b).
%Note that large $\theta_{\rm SH}$ does not cause
%the sign reversal of $v_{\rm DW}$.
%
The parameters for the simulation; $\alpha=0.5$,
$\beta=0.25$~\cite{addenda}, $M_{\rm s}=5.0\times 10^5$ A/m, the
perpendicular magnetic anisotropy constant $K_{\rm u}=1.0\times
10^6$ J/m$^3$, the exchange stiffness constant $A_{\rm
ex}=1.0\times 10^{-11}$ J/m, $P=0.7$, $\gamma/2\pi=28.0113$
GHz$\cdot$T$^{-1}$, and $t_{\rm F}=0.6$ nm.
%
%In these plots, $\lambda$ within the definition of
%$\tilde{\alpha}_R$ is defined as the DW width for $j_e=0$, and the
%DW tilting angle is defined as the value of $\phi$ at the position
%where $m_z=0$.
%
%The simulation used the system size of 2000 nm$\times$200 nm along
%the $x$ and $y$ directions with the cell size of 2 nm$\times$ 200
%nm.
}
\label{Fig:simulation}
\end{figure}
%%%%%%%%%%%%%%%%%%%%%%%%%%%%%%%%%%%%%%%%%

Figures~\ref{Fig:simulation}(a) and (b) show micromagnetic
simulation results of Eq.~(\ref{Eq:LLG-modified}) for the stable
DW structure in Fig.~\ref{Fig:Bloch DW}(a). The inset in
Fig.~\ref{Fig:simulation}(a) shows $v_{\rm DW}$
%of the stable DW configuration [Fig.~\ref{Fig:Bloch DW}(a)]
as a function of $\tilde{\alpha}_{\rm R}$ at fixed $j_e\equiv
\vhat{x}\cdot \vec{j}_e=+3\times 10^{11}$ A/m$^2$ (amounting to
$b_j=+25$ m/s). Note that as $\tilde{\alpha}_{\rm R}$ increases,
$v_{\rm DW}$ changes its sign from negative (along the electron
flow direction) to positive (against the electron flow direction).
The main panel in Fig.~\ref{Fig:simulation}(a) shows $v_{\rm DW}$
as a function of $j_e$ at two fixed values of $\tilde{\alpha}_{\rm
R}$, 0 (black squares) and 10 (red circles). For
$\tilde{\alpha}_{\rm R}=10$, $v_{\rm DW}$ changes from negative to
positive at $j_e\approx 3.5\times 10^{11}$ A/m$^2$
%These results are consistent with our explanation for the anomaly (ii).
%
and goes above +500 m/s for $j_e>1.5\times 10^{12}$ A/m$^2$. Thus
both the anomalies (ii) and (iii) can be explained by RSOC if
$\tilde{\alpha}_{\rm R}$ is sufficiently larger than 1.
By the way, results for the DW structure in Fig.~\ref{Fig:Bloch
DW}(b) are not shown since, when $\tilde{\alpha}_{\rm R}=10$, it
is unstable and switches to the DW structure in
Fig.~\ref{Fig:Bloch DW}(a) for $j_e>7.4\times 10^{10}$ A/m$^2$.
%is stable only at very small current densities ($j_e<7.4\times
%10^{10}$ A/m$^2$) when $\tilde{\alpha}_{\rm R}=10$. For higher
%current densities, the DW structure changes to that in
%Fig.~\ref{Fig:Bloch DW}(a).

Next we demonstrate that the above explanation for the anomalies (ii) and (iii) does not work if
the FL-STT is absent or very small.
%Here we emphasize that both SL- and FL-STTs are required to
%explain the anomalies (ii) and (iii).
%
A recent experiment~\cite{Liu11preprint} on a somewhat different
ultrathin magnetic system Pt/Co/Al reported that the SL-STT has a
sizable magnitude but the FL-STT is negligibly small. This
situation has been attributed~\cite{Liu11preprint} to a {\it
perpendicular} spin current in Co generated from an {\it in-plane}
charge current in the heavy metal layer (Pt) through the SHE.
In the absence of the FL-STT, the stabilities  of the two DW
configurations in Fig.~\ref{Fig:Bloch DW} are governed by the
SL-STT.
%Now the SL-STT is responsible not only for the dynamics
%equivalence breaking but also for the stability equivalence
%breaking of the two DW structures in Figs.~\ref{Fig:Bloch DW}(a)
%and (b).
Recalling that the DW anisotropy energy of the Bloch DW favors
$\phi=0$ and $\pm 180^\circ$, and disfavors $\pm 90^\circ$, it is
evident that the SL-STT tends to destabilize/stabilize the DW
structure in Fig.~\ref{Fig:Bloch DW}(a)/(b). Thus when the
SHE-induced SL-STT is sufficiently strong, the DW structure in
Fig.~\ref{Fig:Bloch DW}(b) is the only stable structure, which is
opposite to the choice made by the FL-STT.
Micromagnetic simulation for this stable DW structure indicates
[inset in Fig.~\ref{Fig:simulation}(c)] that $v_{\rm DW}$ does not
change its sign as $\theta_{\rm SH}$ grows, which is in contrast
to the RSOC effect [inset in Fig.~\ref{Fig:simulation}(a)]. Here
$\theta_{\rm SH}$ is the spin Hall angle representing the strength
of SHE.
%$v_{\rm DW}$ for this stable DW structure is calculated by
%micromagnetic simulation and shown in the inset in
%Fig.~\ref{Fig:simulation}(c) as a function of the spin Hall angle
%$\theta_{\rm SH}$, representing the strength of SHE.  Note that
%there is no sign change, in contrast to the RSOC effect [inset in
%Fig.~\ref{Fig:simulation}(a)].
This result is natural since for the stable DW structure, the
SL-STT is {\it parallel} to the nonadibatic STT at the DW center.
Note also that large enhancement of $|v_{\rm DW}|$ does not occur
[Fig.~\ref{Fig:simulation}(c)] either, since the SL-STT in
Fig.~\ref{Fig:Bloch DW}(b) suppresses the deviation of $\phi$ from
$\pm 180^\circ$, thereby suppressing its own magnitude, which is
proportional to $\sin \phi$. Thus SHE alone cannot explain the
anomalies (ii) and (iii).
By the way, the DW structure in Fig.~\ref{Fig:Bloch DW}(a) may
exhibit the sign-reversed $v_{\rm DW}$ [red circles in
Fig.~\ref{Fig:simulation}(c) near $j_e=4\times 10^{11}$ A/m$^2$]
right before it loses its stability at $j_e\approx 5\times
10^{11}$ A/m$^2$ [downward jump of red circles by $180^\circ$ in
Fig.~\ref{Fig:simulation}(d)]. However this sign-reversal should
be contrasted with the sign-reversal observed~\cite{Miron11NM} in
the current range, where only one stable DW structure exists.

To conclude, we presented the theory of the RSOC-induced STTs,
which explains various anomalous features of the magnetization
dynamics in ultrathin magnetic systems.
It will be valuable for future works to utilize ultrathin magnetic
systems for next generation spintronic devices.
Important future research directions include exploring other
material combinations such as Ta/CoFeB/MgO, utilizing both RSOC
and SHE to achieve enhanced properties, and understanding the
dependence of the RSOC strength on magnetic layer thickness and
material combinations.
It is also interesting to explore possible connections between
this work and other interesting phenomena in ultrathin magnetic
systems such as magnetization control via electric
voltage~\cite{Shiota12NatMat} and strong perpendicular magnetic
anisotropy~\cite{Ikeda11NatMat} in Ta/CoFeB/MgO systems.
Lastly, during the preparation of this manuscript, we received a
calculation~\cite{Manchon11preprint}, which also presents
a derivation of the SL-STT but does not discuss implications of RSOC
on the DW motion.

We gratefully acknowledge M. D. Stiles, R. McMichael, and W. Rippard for valuable
comments. This work was financially supported by the NRF
(2010-0014109, 2010-0023798, 2011-0030784, 2011-0028163) and BK21.
KWK acknowledges the financial support by the NRF (2011-0009278).

%%%%%%%%%%%%%%%%%%%%%%%%%%%%%%%%%%%%%%%%%%%%%%%%%%%%


\begin{thebibliography}{99}
\bibitem{comment:anomalous velocity correction} The correction
to this approximation is ignored since it is not
    a response of an electric current.
\bibitem{comment:algebra}
%$\vec{v}$ is assumed to be spatially homogeneous.
When $\vec{v}_\pm$ is not homogeneous, the spin diffusion dynamics
modifies the relation between the spin current and the electric
field~\cite{Kim11PRB} but it does not modify the RSOC-induced STTs
expressed in terms of the current density.
\bibitem{addenda} $\beta$ is assumed to be smaller than $\alpha$ to be
    consistent with experiments~\cite{Miron11Nat,Miron10NM,Miron11NM};
    $\alpha$ is about 0.5 ~\cite{Miron10NM}. We estimate $\beta$ to be of
    order 0.1 in the experiments since the SL-STT~\cite{Miron11Nat} is about
    one order of magnitude smaller than the FL-STT~\cite{Miron10NM} at the
    same current density. By the way, estimation of $\beta/\alpha$ from
    $v_{\rm DW}$ is incorrect since RSOC strongly modifies $v_{\rm DW}$.
\end{thebibliography}
\end{document}